\documentclass[aps,pra,onecolumn,superscriptaddress,nofootinbib,longbibliography]{revtex4-1}
\usepackage{times}
\usepackage{inconsolata}

\usepackage{amsmath}
\usepackage{amssymb}
\usepackage{tabularx}

\usepackage[hidelinks]{hyperref}
\usepackage{graphicx}
\usepackage[utf8]{inputenc}

\usepackage{enumitem}
\setlist{noitemsep}
\setlist{nolistsep}

\def\R{\mathbb R}

\begin{document}

\title{{Theano}: A {Python} framework for fast computation of mathematical expressions}

\collaboration{The Theano Development Team}
\email{theano-dev@googlegroups.com}
\homepage{http://deeplearning.net/software/theano}
\thanks{code available at \url{https://github.com/Theano}}


\affiliation{Montreal Institute for Learning Algorithms (MILA), Université de Montréal, QC, Canada}
\affiliation{École Polytechnique de Montréal, QC, Canada}
\affiliation{CIFAR Senior Fellow}
\affiliation{CIFAR Fellow}
\affiliation{CIFAR Associate Fellow}

\author{Rami~Al-Rfou}
\affiliation{Stony Brook University, NY, USA}
\author{Guillaume~Alain}
\author{Amjad~Almahairi}
\affiliation{Montreal Institute for Learning Algorithms (MILA), Université de Montréal, QC, Canada}
\author{Christof~Angermueller}
\affiliation{University of Cambridge, UK}
\affiliation{European Bioinformatics Institute, European Molecular Biology Laboratory, Cambridge, UK}
\author{Dzmitry~Bahdanau}
\author{Nicolas~Ballas}
\author{Frédéric~Bastien}
\affiliation{Montreal Institute for Learning Algorithms (MILA), Université de Montréal, QC, Canada}
\author{Justin~Bayer}
\noaffiliation
\author{Anatoly~Belikov}
\affiliation{Bauman Moscow State Technical University, Russia}
\author{Alexander~Belopolsky}
\affiliation{Enlightenment Research LLC, New York, NY, USA}
% ...

\begin{abstract}

Theano is a Python library that allows to define, optimize, and evaluate mathematical expressions involving multi-dimensional arrays efficiently. Since its introduction in~\cite{bergstra+al:2010-scipy} it has been one of the most used CPU and GPU mathematical compilers -- especially in the machine learning community~\cite{bergstra+all-Theano-NIPS2011} -- and has shown steady performance improvements~\cite{Bastien-Theano-2012}.
Theano is being actively and continuously developed since 2008, multiple frameworks have been built on top of it and it has been used to produce many state-of-the-art machine learning models.

The present article is structured as follows.
Section~\ref{sec:overview} provides an overview of the Theano software and its community.
Section~\ref{sec:main} presents the principal features of Theano and how to use them, and compares them with other similar projects. Section~\ref{sec:new} focuses on recently-introduced functionalities and improvements.
Section~\ref{sec:benchmarks} compares the performance of Theano against Torch7~\cite{Torch-2011} and TensorFlow~\cite{tensorflow2015-whitepaper} on several machine learning models.
Section~\ref{sec:challenges} discusses current limitations of Theano and potential ways of improving it.

\end{abstract}

\maketitle

\section{Overview}
\label{sec:overview}

    \subsection{Vision}
    \label{sec:overview:vision}

Theano allows a user to symbolically define mathematical expressions and have them compiled in a highly optimized fashion either on CPUs or GPUs (the latter using CUDA)\footnote{Some OpenCL support is available in the new GPU back-end, but it is still limited and experimental.}, just by modifying a configuration flag.
Furthermore, Theano can automatically compute symbolic differentiation of complex expressions, ignore the variables that are not required to compute the final output, reuse partial results to avoid redundant computations, apply mathematical simplifications, compute operations in place when possible to minimize the memory usage, and apply numerical stability optimization to overcome or minimize the error due to hardware approximations.
To achieve this, the mathematical expressions defined by the user are stored as a graph of variables and operations, that is pruned and optimized at compilation time.

The interface to Theano is Python, a powerful and flexible language that allows for rapid prototyping and provides a fast and easy way to interact with the data. The downside of Python is its interpreter, that is in many cases a poor engine for executing mathematical calculations both in terms of memory usage and speed. Theano overcomes this limitation, by exploiting the compactness and ductility of the Python language and combining them with a fast and optimized computation engine.

Theano's API mimics NumPy~\cite{numpy, scipy}, a widely adopted Python library that provides an n-dimensional array data type and many functions for indexing, reshaping, and performing elementary computations (exp, log, sin, etc.) on entire arrays at once.
This allows Python users to rapidly switch to Theano using a familiar syntax and set of instructions -- extended with advanced features, such as automatic gradient computation, numerical stability improvements and optimization -- and generate a high-performance code for CPU as well as for GPU, without requiring changes to the user code.
Theano has also been designed for easy and fast extensibility through the definition of custom graph expressions written in Python, C++, or CUDA.

    \subsection{Community}
    \label{sec:overview:community}

Theano is a free, open-source software, licensed under the New (3-clause) BSD license. It relies on a wide and very active community of developers and users worldwide.

The main communication channels with the developers are the project's GitHub page\footnote{\url{https://github.com/Theano/Theano/}} for bug reports, feature requests, and pull requests, and the theano-dev mailing list,\footnote{\url{https://groups.google.com/group/theano-dev/}} which has 675 subscribers. Support for users is provided by the community at theano-users\footnote{\url{https://groups.google.com/group/theano-users/}} (more than 3000 members) and on StackOverflow\footnote{\url{http://stackoverflow.com/questions/tagged/theano}} (more than 1000 questions asked).
 PyPI\footnote{\url{https://pypi.python.org/pypi}} counted 38k downloads of Theano packages during the last month.

Since the project development migrated to GitHub in 2011, Theano has been forked 1280 times. Around 250 developers have actively contributed to the code base, and numerous others have played a role in the community, asking, answering or curating questions, helping discussing the development needs, and writing documentation, tutorials,\footnote{For instance, the deep learning tutorials at \url{http://deeplearning.net/tutorial/}} or even full-fledged software projects based on Theano.

    \subsection{Software based on Theano}
    \label{sec:overview:software}

Several software packages have been developed to build on the strengths of Theano, with a higher-level user interface, more suitable for certain goals. For instance, machine learning and deep learning packages, such as
Pylearn2~\cite{pylearn2_arxiv_2013},
Blocks~\cite{Van+al-arxiv-2015},
Lasagne~\cite{sander_dieleman_2015_27878},
and Keras~\cite{chollet2015keras},
have been developed with the goal of making it easier to express the architecture of deep learning models, and training algorithms, as mathematical expressions to be evaluated by Theano.

Another example is PyMC3~\cite{SalvatierWF16}, a probabilistic programming framework that uses Theano to derive expressions for gradients automatically, and to generate C code for fast execution.

\section{Main features}
\label{sec:main}

Theano defines a \emph{language} to represent mathematical expressions and manipulate them (Section~\ref{sec:main:language}),
a \emph{compiler} to create functions that can compute values for these expressions (Section~\ref{sec:main:compiler}),
and a \emph{library} which will execute these functions when evaluated on numeric values (Section~\ref{sec:main:library}).
We also explain how Theano can be extended (Section~\ref{sec:main:extending}).
Finally, we provide some comparison points with related software (Section~\ref{sec:main:comparison}).

    \subsection{Mathematical expressions}
    \label{sec:main:language}

        \subsubsection{Graph structure}
        \label{sec:main:language:graph}

Theano represents symbolic mathematical expressions as directed, acyclic graphs. These graphs are also bipartite, containing two kinds of nodes:
\begin{itemize}
    \item \textbf{Variable} nodes (or variables), which represent \emph{data}, usually tensors;
    \item \textbf{Apply} nodes, which represent the application of \emph{mathematical operations}.
\end{itemize}
In practice, variables are used for graph inputs and outputs, as well as for intermediate values. During the execution phase, values will be provided for input variables, and computed for intermediate and output ones. An Apply node has inputs and outputs, which are Variable nodes; it represents the application of a mathematical operation (or Op) on its input variables. A Variable node can be the input to several Apply nodes, but can be the output of at most one (graph inputs are not the result of any computation). This corresponds to the single static assignment (SSA) form in compiler design, in that a variable is the result of only one assignation.

This structure is similar to dataflow graphs~\cite{arvind1986dataflow}, where Apply nodes would correspond to operations nodes (the only kind of nodes), and Variable nodes would correspond to arcs in the dataflow graph. The main difference is that a single intermediate Variable node can be an input to several Apply nodes, whereas a dataflow graph would require different arcs, one for each of the next operations.

Variables are strongly typed, they enforce some conditions on the values that can be associated with them. These types are known since the construction of the graph. The main categories of types are:
\begin{itemize}
    \item \verb|TensorType|, which represents n-dimensional arrays in the main memory, the values associated with variables of that type are NumPy \verb|ndarray| objects;
    \item \verb|CudaNdarrayType|, which represents n-dimensional arrays in GPU memory, associated with \verb|CudaNdarray| objects, used in the legacy GPU back-end;
    \item \verb|GpuArrayType|, associated with \verb|GpuArray| objects, its equivalent in the new GPU back-end;
    \item \verb|Sparse|, for main-memory sparse matrices, represented by SciPy CSC or CSR matrices.
\end{itemize}
The number of dimensions and the data type (float32, int64, etc.) are part of the type, as well as what we call the \emph{broadcastable pattern}, which indicates which dimensions are guaranteed to have a shape of 1. Otherwise, the shape is not part of the type, and neither is the memory layout (strides).

        \subsubsection{Building a graph}
        \label{sec:main:language:build}
        
A computation graph is usually constructed by creating free symbolic variables first, corresponding to the inputs of the graph. Since variables are strongly typed in Theano, the type of these variables has to be specified at creation time. By calling Python functions on variables, the user can then interact with them in a direct and natural way. This is reflected under the hood by the creation of Apply nodes and new Variable nodes that extend the graph. The \verb|tensor| module exposes many of the functions provided by NumPy for tensor operations, to present a familiar interface to users. Some of these add a single Apply node and its output to the graph, returning the output Variable node, while other build more complex graphs with Apply nodes corresponding to different Ops, combined in such a way that the returned variable represents the expected result.

It is also possible to clone an existing graph, or a part of it. In that case, what was an intermediate variable in the original graph could become a free input, or an output, of the cloned graph. It is also possible to clone with replacements, which make it possible to plug together different disconnected graphs, making inputs into intermediate Variable nodes.

        \subsubsection{Symbolic differentiation}
        \label{sec:main:language:grad}

A useful way of deriving gradients is by applying the chain rule backwards through the graph, from a scalar cost towards the inputs (or parameters). This procedure is known as gradient back-propagation, or as the backward or reverse mode of differentiation. For instance, if we have three functions $f: \R^M \rightarrow \R$, $g: \R^N \rightarrow \R^M$, and $C: \R^N \rightarrow \R$ so that $C(x) = f(g(x))$, then:
\[
    \left.\frac{\partial C}{\partial x}\right|_x =
              \left.\frac{\partial f}{\partial g}\right|_{g(x)}
              \cdot \left.\frac{\partial g}{\partial x}\right|_x
\]
Instead of computing (and storing in memory) explicitly the whole $M \times N$ Jacobian matrix, $\left.\frac{\partial g}{\partial x}\right|_x$, all we need is a function
$\nabla g_x: \R^M \rightarrow \R^N, v \mapsto v \cdot \left.\frac{\partial g}{\partial x}\right|_x$ that computes the vector-Jacobian dot product for any vector $v$. This can be generalized easily to functions with several inputs, which can be multi-dimensional arrays.

Most of Theano Ops implement a \verb|grad| method that, given symbolic variables for $x$ and $v$, will return a symbolic expression of $\nabla g_x(v)$, where $g$ is the function represented by that Op. \verb|theano.grad| traverses the graph following the usual back-propagation algorithm, calling the \verb|grad| method on each Apply node's Op, passing that node's input as $x$ and the gradient coming from the subsequent operations as $v$. This builds a symbolic expression for the gradient of the cost with respect to variables. These gradients are symbolic variables that are part of the graph as well, so it is possible to use them as parts of other symbolic expressions (to express a learning rule, for instance), and even to traverse the graph again to obtain higher-order derivatives.

Many Theano Ops also implement an \verb|R_op| method, computing a symbolic expression for the the Jacobian-vector dot product, $R g_x: {\R^N \rightarrow \R^M}, {v \mapsto \left.\frac{\partial g}{\partial x}\right|_x \cdot v}$. This is the R-operator introduced by~\cite{pearlmutter94}, and corresponds to the forward mode of differentiation. \verb|theano.Rop| traverses the graph from inputs to outputs, calling the \verb|R_op| method on each Apply node's Op.

        \subsubsection{Scan: Symbolic loops}
        \label{sec:main:language:scan}

Since the computation graph is acyclic, and its structure is fixed and independent from the actual data, it can be a challenge to express loops symbolically. One option, when the number of steps in the loop is fixed, is to explicitly unroll the loop, adding to the computation graph the computation of each of the iterations multiple times. Unfortunately, this makes it impossible to iterate over a sequence of unknown length, or to iterate a variable number of times depending on the value of the data.

To sidestep these issues, Theano implements a special Op called \emph{Scan}, which abstracts the entire loop in a single Apply node in the graph. That single node contains a full computation graph, isolated from the main one, that represents the computation done during each iteration of the loop. The scan node handles the communication between the \emph{external} or \emph{outer} computation graph it belongs to, and the \emph{internal} or \emph{inner} graph. It is also responsible to manage the bookkeeping between the different iterations.

The gradient of a Scan operation is implemented as another Scan operation, which iterates over reversed sequences, computing the same gradient as if the loop had been unrolled, implementing what is known as \emph{back-propagation through time}. Similarly, the R operator is also a Scan operation that goes through the loop in the same order as the original Scan.

    \subsection{The compilation phase}
    \label{sec:main:compiler}

The compilation phase produces a Theano function (a Python callable object) able to compute values for specified \emph{output} symbolic variables, given values for \emph{input} variables. The set of input and output variables have to be provided when compiling the function, but the inputs do not have to be inputs to the full computation graph, and outputs do not have to be ultimate outputs either. It is possible to compile a function going from some intermediate variables of the graph to other intermediate variables, as long as the set of inputs contains all the information to compute the set of outputs. Several Theano functions can be compiled, computing different parts of the same computation graph.

During the compilation of a Theano function, first the relevant portion of the computation graph is cloned, then it gets rewritten by the application of \emph{graph optimizations}, next some optimized C++ or CUDA code gets generated and compiled if necessary, and finally a callable object is built and returned to the user.

        \subsubsection{Graph optimizations}
        \label{sec:main:compiler:opt}

The computation graph structure makes it possible to replace parts of the graph. For instance, a Variable node which is the output of one particular Apply node could be replaced by the output of a different Apply node, as long as they have the same type. Optimizations specify how to perform replacements of variables by other variables representing an equivalent computation. Some of them are \emph{local}, which means they only look at one Apply node and can replace its outputs, some of them are \emph{global}, and can examine the whole computation graph and perform arbitrary substitutions. Optimizations are mostly organized into the stages described below, even if there is some overlap.
\begin{itemize}
    \item \emph{Canonicalize:} Put the graph in a canonical form, to ease the task of subsequent optimizations (for instance, $x*x \Rightarrow x^2$). It performs some simplifications as well, like removing duplicate computations, removing some unnecessary computations ($xy/y \Rightarrow x$), and computing the value of expressions if all their inputs are known (constant-folding, $2 + 2 \Rightarrow 4$).
    \item \emph{Stabilize:} Increase numerical stability, for instance $\log{(1+x)} \Rightarrow \mathrm{log1p}(x)$, where log1p is a stable implementation for small $x$.
    \item \emph{Specialize:} Insert faster implementations of operations. For instance, successive element-wise operations are fused together to avoid having to loop over a tensor several times.
    \item \emph{GPU:} Replace the default version of Ops and variables by GPU-specific versions, using either the old or new back-end, if a GPU is requested. Transfer Ops (CPU-to-GPU or GPU-to-CPU) are inserted so that the type of inputs and outputs is preserved, and around CPU-only operations.
    \item \emph{Inplace:} Replace the default version of Ops by a version that can work in-place, as a view or destructive operation over its inputs.
    The array types used by Theano, like \verb|ndarray|, support arbitrarily-strided arrays, so all transposition operations, as well as basic slicing, can happen in place, in constant time.
    Some operations, like most element-wise ones, can overwrite their input and return it, to avoid allocating memory. Since destructive operations introduce additional dependencies between Apply nodes (a value can only be overwritten by the \emph{last} operation to read it), dependency cycles have to be detected and prevented.
    \item \emph{Scan:} Optimize performance and memory use of Scan nodes. For instance, only keep the value for the last step of an output in memory if the whole sequence is not needed, merge different Scan nodes to perform computations only once, and move invariants out of the loop.
\end{itemize}

While individual optimizations or groups of optimizations can be individually enabled or disabled, some optimizers (sets of optimizations) are predefined: \verb|'None'| does not include any optimization, \verb|'fast_compile'| includes only canonicalization and transfer to the GPU, and \verb|'fast_run'| (the default) includes most optimizations except for experimental and ``unsafe'' ones (removing assertions).

        \subsubsection{Shared variables}
        \label{sec:main:compiler:shared}

Shared variables are symbolic variables that are associated with persistent values, that are shared between Theano functions. They can only be input variables (not intermediate ones), since their value is not the result of the computation of an Apply node. Shared variables are implicit inputs to all the Theano functions using them.

When compiling a Theano function, it is possible to specify \emph{update expressions} for shared variables. These expressions are symbolic variables that represent the new value to assign the the shared variables at the end of each function execution. They are implicit outputs of the function, and will be computed along with the other outputs, before the value gets updated. Such update rules make it possible to update the array in-place in some cases, rather than returning a different array.

It is also possible to explicitly assign a new value to an existing shared variable, outside of a Theano function, as long as it is compatible with its type. Since the shape is not part of the type, it is possible for the shape of a shared variable to change. If a GPU is enabled, shared variables will be created on the GPU by default, to avoid transfers (this only works for \verb|float32| arrays in the old back-end).

        \subsubsection{C code compilation and caching}
        \label{sec:main:compiler:c_code}

The code to compute output values given input values for each Op can be implemented either in Python or in C++ (or CUDA for GPU Ops), using the C API from Python and NumPy (and from CudaNdarray or GpuArray for GPU).

After the function graph is optimized, each Op generates the C++ or CUDA code for a Python module implementing that computation (including reading and writing from the right storage map), which is then compiled, and imported.

A persistent cache on disk makes it possible to avoid generating code twice for the same Op, and to avoid compiling again when different Ops generate the same code (this can happen for the same operation applied on different data types, or different numbers of dimensions, for instance).

    \subsection{Function execution}
    \label{sec:main:library}

Theano includes a runtime engine that, upon a Theano function call, determines the computation to be executed on which data and in what order, and orchestrate their evaluation.
This was originally done by forward-traversing graphs from input to output, requiring all branches to be evaluated before outputs could be returned. The default runtime now uses a virtual machine (VM) system. By running small code units (each corresponding to an Apply node for one Op) and ignoring branches not necessary for correct computations, lazy evaluation is now possible.

The runtime uses a data structure containing pointers to storage for each variable (inputs and outputs of each Apply node), ordering constraints, pointers to the functions performing the computations, and information on what has been computed and needs to be computed in the current call. If the speed of execution is more important than memory usage, it is possible to keep references to ndarrays containing intermediate results, to prevent Python's garbage collection from freeing them, and to re-use it for the next run of the function, through the configuration flag \verb|allow_gc=False|. The default is to allow the garbage collector to free the storage of intermediate values.

The C implementation of that VM (CVM) is the default runtime. Not only does this increase performance by running the runtime loop in C, if a C implementation of an Op is available, the CVM can directly execute it. This eliminates the overhead from a Python function call, which is especially advantageous when performing many operations on small operands.

A Python implementation is also available. It is more flexible and easier to instrument, which is useful to collect more profiling information (for instance, memory usage) and add callbacks for debugging.

    \subsection{Extending Theano}
    \label{sec:main:extending}

If the existing Theano library does not include the operations required for a particular model, the framework was designed for easy extensibility. New Ops can be written by specifying the type of their input and output variables, and providing Python code to perform the evaluation. That Python code can use bindings to external high-performance libraries, or Cython, for instance. Methods can also be added to specify expressions for gradients and the  R-operator (see Section~\ref{sec:main:language:grad}), and shape inference. Theano's self-testing functions can be used to validate outputs and check symbolic gradients against numeric evaluations among others.

As mentioned above, operators can also be implemented directly in C++ or CUDA. The raw code can be supplied as a string that the Python code uses to produce the code used by the graph compiler. 
For added convenience, Theano can now load code from an external C-like file with the \verb|COp| class.
The file is divided into sections that map to the different pieces of code that Theano requires.
Keeping the Python and C code separate allows more readable code with better indentation. 
It also enables a clearer view of the C code itself since you can use your favorite C editor to modify that file with syntax highlighting.

A user can then write a new \emph{optimization} to automatically insert that optimized operation in the computation graph, instead of the more naïve or slow version. This is especially useful when implementing an operation on GPU.

    \subsection{Related software}
    \label{sec:main:comparison}
Although Theano is developed and mainly used for research in machine learning and deep learning, it is not a deep learning framework in itself (see Section~\ref{sec:overview:software} for some machine learning frameworks based on Theano). However, it makes sense to compare the core features of such systems with Theano, as they all support the definition of a mathematical model in a symbolic way, and implement some automatic gradient computation.

TensorFlow~\cite{tensorflow2015-whitepaper} has a core in C++ and includes most of the features from Theano, in particular the graph-compiling approach, and symbolic differentiation (on full layers as well as on elementary operations), all directly accessible from Python through the API. In addition, it has a focus on distributed, multi-node computation. Even though a graph-rewriting engine is present (and used to distribute computation across devices, for instance) it does not seem to be used for mathematical expressions simplification or kernel fusion at the moment.

Torch7~\cite{Torch-2011} has a different approach: it implements efficient CPU and GPU computation kernels in C and makes them available in Lua, but does not provide gradient expressions for elementary operations. Instead, packages like `nn` and `cunn` feature higher-level \emph{layers} that can store parameters and provide methods to compute values for forward propagation, gradient back-propagation, and parameter updates.
Many packages extend Torch's features, in particular Autograd\footnote{\url{https://github.com/twitter/torch-autograd/}} provides automatic differentiation of code written in Torch, by building a graph that records the evaluation of expressions (even through loops and conditionals), and playing those records back to build an expression graph for gradients. That graph is symbolic as well, making it possible to express higher-order gradients. Moreover, an optimizer can rewrite the graph to make it more efficient to evaluate.

MXNet~\cite{mxnet-2015-arxiv} and Caffe~\cite{jia2014caffe}, both written in C++, feature the same kind of higher-level layers as Torch. MXNet can also express the gradients through those layers as symbolic layers themselves, giving more flexibility for the dispatching of the computation to different devices, and for memory reuse. It also allows distributed computation over multiple nodes. Caffe2\footnote{\url{https://github.com/Yangqing/caffe2}} is an experimental rewrite of Caffe that features explicit symbolic gradients in the computation graph, rather than a ``backward'' method of the layers.

Neon\footnote{\url{http://neon.nervanasys.com/}} and Chainer~\cite{chainer_learningsys2015} are two other machine learning frameworks written in Python, with GPU kernels, that feature symbolic computation graphs and symbolic differentiation. Neon's most prominent feature is its collection of highly-optimized GPU kernels, in particular for operations used in neural networks. Chainer instead builds its computation graph dynamically at the same time as its first evaluation, making it easier to express loops and conditionals.

\section{New features}
\label{sec:new}

Over the last couple of years, multiple improvements have been made in Theano, in particular for faster execution, including support for more operations on the GPU and multiple-GPU support (Section~\ref{sec:new:performance}), faster graph optimization, especially for larger graphs (Section~\ref{sec:new:opt}), and ease of use,  with better error messages and tools for introspection, visualization, and debugging (Section~\ref{sec:new:debugging}).

    \subsection{Increased performance}
    \label{sec:new:performance}

        \subsubsection{Abstract Ops and 2D convolutions}
        \label{sec:new:performance:abstractconv}
Convolution operations are at the core of Convolutional Neural Networks (CNNs) that have lead to spectacular advances in machine learning problem involving visual data~\cite{krizhevsky2012imagenet}.
A more detailed description of the convolution operations can be found in~\cite{dumoulin-visin:2016arXiv}.

The multiplication of available implementations for convolution (CPU-GEMM, GPU-cuDNN, GPU-GEMM, FFT, \ldots{}) available in Theano  has increased the need of a flexible convolution interface that easily allows to switch between those implementations, each implementation having different speed and memory trade-off, as well as different software dependencies.
To suit this need, Theano 0.8 introduces abstract Ops that allows to disentangle the interface of an Op to their actual implementation.
An abstract Op introduces is a place-holder Apply node in the graph, corresponding to a given operation, that does not provide an actual implementation. For each optimized implementation of that operation, there is an optimization that will insert an Apply node for that optimized Op instead of the abstract Apply node during the compilation phase.

In particular, Theano proposes three abstract Ops for convolution: \verb|AbstractConv2d|, \verb|AbstractConv2d_gradInputs|, and \verb|AbstractConv2d_gradWeights|, that correspond respectively to the forward convolution, the convolution gradient w.r.t.\ inputs and the convolution gradient w.r.t.\ weights. Each abstract Op can be replaced by one of the different implementations.
By default, if a GPU is enabled and cuDNN is available, Theano will use it (see Section~\ref{sec:new:performance:cudnn}), otherwise it will fall back to using the GEMM version. A slow, Python-only implementation is part of the abstract Ops for debugging purposes. The optimizations can be included or excluded using the configuration flags, which makes it possible to manually select a specific convolution implementation.

        \subsubsection{Using cuDNN}
        \label{sec:new:performance:cudnn}

Efficient CUDA primitives for neural networks are implemented in the cuDNN library~\cite{cudnn-2014-arxiv}, in particular convolutions, pooling, and their gradients. Several implementation of convolutions (and gradients) are provided, with the same interface, with performance and memory usage that depends on the actual shape of the data and filters. Since the best implementation can be different for different convolutions in the same model (depending on their size) and on different hardware (depending on the available memory), cuDNN also provides a heuristic to guess the best algorithm given shapes, and to actually time the different implementations (that are feasible given the available free memory) and select the fastest one.

Theano wraps cuDNN 2D and 3D convolutions and their gradients, and provide options to select the algorithm to use, either explicitly or using one of the following special values: \verb|'guess_once'|, \verb|'guess_on_shape_change'|, \verb|'time_once'|, or \verb|'time_on_shape_change'|. This selection can be done individually for each Apply node in the graph, and configuration flags select the global default for the forward convolution, the gradient w.r.t. the data, and the gradient w.r.t. the weights. Theano also wraps pooling operations, as well as softmax and log-softmax operations. More operations will be added in the future.

        \subsubsection{CNMeM integration}
        \label{sec:new:performance:cnmem}

Another improvement to the GPU performance comes integrating the CNMeM library,\footnote{The original code is available at \url{https://github.com/NVIDIA/cnmem}, Theano includes a copy of it.} and using the allocator and deallocator it provides. The main issue was that calling \verb|cudaFree| is synchronous, so it forces the synchronization of all the streams on the device, waiting for them to finish, which seriously limited the potential for parallel execution of different kernels. A previous option was to keep memory allocated for intermediate values between calls, as mentioned in Section~\ref{sec:main:library}, but the amount of memory typically available on GPU devices is limited.

CNMeM works by allocating large memory pools using \verb|cudaMalloc|, returning chunks of it when its allocator is called, and keeping track of which ones are released by its deallocator. Theano makes it possible to reserve part of the GPU memory from the start, using \verb|lib.cnmem=0.9| to reserve 90\% of the memory for CNMeM.
The new GPU back-end does not use CNMeM, but implements a similar strategy, with asynchronous allocator and deallocator and a memory pool.

        \subsubsection{Improvements in Scan}
        \label{sec:new:performance:scan}

Important speed improvements have been made to Scan, in addition to making it more stable, and supporting more cases. The time to optimize and compile graphs containing Scan Apply nodes has been reduced a lot, and the execution time of the resulting function has improved as well.

The optimizations related to Scan (pushing computation out of the loop, removing useless computation) have been improved so they can be applied faster. Additional optimizations have been added, so that more computation can be moved out of the loop, for increased execution speed.

The execution back-end of Scan has been made more efficient as well, by removing some of the bookkeeping overhead, and making the internal function write directly into the right output buffer at each execution step, rather than having to copy the intermediate results each time.

The \verb|grad| method of Scan has been rewritten to scale better in the case of large numbers of input and output variables, and to generate a cleaner graph. That cleaner graph can lead to a faster optimization time, since less rewriting is needed and the inner graph is smaller, and faster execution as well. In the case of nested symbolic loops, the observed speed up in compilation time was sometimes huge, going from hours to minutes.

Finally, an additional keyword, \verb|strict|, has been added to the \verb|scan| function. It prevents shared variables from being implicitly added as non-sequence inputs to the inner function.
This forces the user to explicitly provide all non-sequences needed in the inner function, which may not be the shared variables themselves, but rather outputs of some computation done of them. In that case, doing so prevents pulling that computation inside the loop, which can speed up the optimization as well as the execution.

        \subsubsection{New gpuarray-based back-end}
        \label{sec:new:performance:newbackend}

Theano now features a new GPU backend based on libgpuarray~\cite{bastien+all-NIPS2011}.
This new back-end brings in several improvements over the previous one.
The most visible improvement is that it supports all the usual data types, instead of being limited to float32 data. In particular, it supports half-precision floating point values (float16).
As did the previous back-end, this one supports views and strides to avoid copies and reuse memory whenever possible.

libgpuarray\footnote{\url{http://deeplearning.net/software/libgpuarray/}, code available at \url{https://github.com/Theano/libgpuarray}} is a separate project with the aim of providing a ndarray-like object on the GPU.
It has a C interface so that it can be reused in other projects that don't use Python. It also supports 64-bit indexing, so that arrays with more than $2^{32}$ elements are supported.

Another noticeable improvement is that we have basic support for OpenCL,
however a sizable portion of the GPU Ops in Theano do not currently support it.
This could be fixed with some porting effort.

The new back-end also allows using multiple GPUs in the same function to do model parallelism.
One example of such a model is the two-stack variant of AlexNet~\cite{krizhevsky2012imagenet}.
This however may be hampered by the Python Global Interpreter Lock (GIL) in some cases, meaning that one will get correct results, but may lose parallelism.

Several new features that help performance are present, but not obvious.
One of these is that all computations are transparently asynchronous, which allows the CPU part of the Ops to execute in parallel with the GPU part.
There is a mechanism keeping track of the dependencies between operations to ensure that the right data is always used.
Data transfers are automatically done on a separate stream, so they can overlap with the computation.

The new back-end is now fully functional, and well tested for correctness. It supports almost all the operations of the old back-end on CUDA-capable devices, including wrapping cuDNN for efficient convolutions, but we are still in the process of tuning some of its kernels for a better performance. In particular, int64-based indexing can be significantly slower than int32, so some adjustments have to be made.

        \subsubsection{Data parallelism with Platoon}
        \label{sec:new:performance:platoon}

To take advantage of multiple computing devices, there are two main approaches: model parallelism and data parallelism.
Model parallelism consists in splitting the model itself into multiple parts and have those parts computed by different devices.
It requires a careful balancing of the size of the parts and of the communication costs to ensure optimal performance.
Data parallelism on the other hand is about splitting your input data in multiple parts, and running multiple copies of the model.
It requires attention to model synchronization so that the copies don't drift apart too much during training, and to the way of aggregating the results produced.

Usually, data parallelism on a single machine is done using multiple threads, but this approach is unworkable in Python because of the Python GIL.
Because of this, we have to turn to multiple processes and this presents a new set of challenges. Platoon\footnote{\url{https://github.com/mila-udem/platoon}} is a package that has been developed to to address those challenges and help train Theano models faster by using data parallelism.

Platoon features a central controller process, that communicates with different worker processes, each using Theano to train a copy of the model on a CPU or GPU. It uses shared memory to share model parameters between workers, in order to avoid inter-process communication overhead.
The communications with the central controller are sent asynchronously, so that the worker does not have to wait for a reply.
There is also a script to launch all the workers and monitor them while running that provides a central ``job'' to wait for on clusters.

Two ways of performing the updates on the central parameters are currently implemented: Asynchronous SGD (ASGD), similar to Downpour SGD~\cite{dean2012large}, and Elastic Averaging SGD (EASGD)~\cite{zhang2015easgd}. Other algorithms can be added by implementing additional parameter synchronization rules.

    \subsection{Faster compilation of graphs}
    \label{sec:new:opt}

        \subsubsection{Faster, simpler optimizer}
        
As mentioned in Section~\ref{sec:main:compiler:opt}, some sets of optimizations are pre-defined and can be easily specified. One of these optimizers, \verb|'fast_compile'|, has recently been upgraded to include the optimizations that transfer computation to a GPU, as well as the optimizations necessary to make those optimizations apply. This drastically shortens the graph optimization time, at the cost of a slightly slower execution time and increased memory usage. That option can speed up the development or prototyping phase of a model, allowing the developer to iterate faster.

        \subsubsection{Swapping updates without recompiling}

It is now possible to copy functions using the \verb!function.copy()! method. This can be useful when creating functions that are similar but use different shared variables or update parameters, for instance when creating test and validation functions. Most importantly, the optimized graph of the original function is copied, meaning compilation only occurs once.

The interface for \verb!copy! lets users specify which shared variables to swap, and whether or not updates are carried over. It is also possible to have copied functions share intermediate storage in memory (storage that is not input or output). When this is combined with disabled garbage collection, this can increase execution speed and save memory.

        \subsubsection{Save and reload optimized graphs}
        
Optimized computation graphs, such as the ones in Theano functions, can now be serialized using the \verb|pickle| module, and get de-serialized without being optimized again. It is possible to force the re-optimization, for instance if the set of optional dependencies available has changed between saving and reloading, in which case the function may not run (if a dependency has been removed) or be sub-optimal (if one has been added). This is especially useful when check-pointing and restoring running experiments. Note that the C++ or CUDA code may still need to be recompiled.

    \subsection{Visualization, debugging, and diagnostic tools}
    \label{sec:new:debugging}

Since the definition of Theano functions is separate from their execution, some specific tools have been developed to help users visualize parts or the whole of the computation graph, pinpoint the origin of errors, and understand what is happening at execution time.
        
        \subsubsection{Interactive visualization with d3viz}
        \label{sec:new:debugging:d3viz}

\begin{figure}[htp]
\centering
\includegraphics[width=0.75\linewidth]{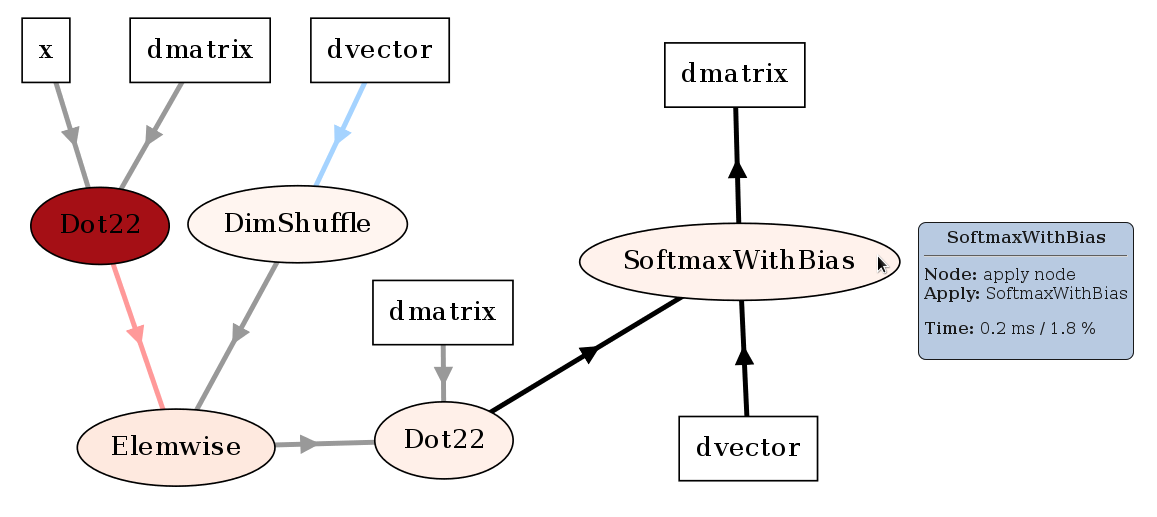}
\caption{Interactive graph visualization with d3viz. Profiling colors have been activated, with redder nodes corresponding to longer computation times. Blue arrows indicate a node returns a view of the input, red arrows indicate a destroyed input.}
\label{fig:d3viz}
\end{figure}

Interactive visualization of computation graphs is now possible with the \verb!d3viz! module, which extends Theano’s printing module. Instead of outputting a text representation (like \verb|debugprint|) or creating a static picture (like \verb|pydotprint|), it creates an HTML file, which can be opened with current web browsers. An example is shown in Figure~\ref{fig:d3viz}.

Several features are supported. Users can zoom different regions, move graphs via drag and drop, and position nodes both manually and automatically. The visualisation can retrieve additional information about nodes and edges such as their data type or definition in the source code, edit node labels and visualize profiling information.
Nested graphs such as \verb|OpFromGraph| nodes can also be explored by expanding or shrinking the nodes as needed.

Internally, \verb!d3viz! represents a compute graph in the Graphviz DOT language, using the pydot package, and defines a front-end based on the d3.js library to visualize it. However, any other Graphviz front-end can be used, which allows to export graphs to different formats such as PNG and PDF.

        \subsubsection{Test values}
        \label{sec:new:debugging:test_values}

Detecting errors in the way a mathematical expression is implemented in Theano can be a challenge, since it is not possible to directly map an intermediate Variable node to the value that will be associated to it at execution time. To mitigate this problem, it is possible to associate a \emph{test value} to input variables, and to compute automatically values associated to intermediate variables as soon as they are defined. This makes it much easier to detect shape mismatches, for instance, or unexpected values.

Note that these values are computed only once, when the graph is built. That means that stability optimizations will not be applied to these values, so NaN (not-a-number) values could be produced during that phase, even if they would not be present when evaluating the optimized graph.

        \subsubsection{NanGuardMode}
        \label{sec:new:debugging:nan}

A frequent symptom of issues when optimizing a model is the appearance of NaN (not-a-number), infinity, or very large values. They can indicate a wide range of issues, e.g., use of un-initialized memory, lack of numerical stability in the computation, divergence of the algorithm itself.

To help diagnosing the appearance of such values, NanGuardMode is an instrumented version of the runtime environment that can check the values of inputs and outputs of each Apply node during execution, and raise an error when some problematic values are detected.

        \subsubsection{The PdbBreakPoint Op}
        \label{sec:new:debugging:breakpoint}

\verb|PdbBreakPoint| is an Op designed to check the value of a \emph{condition}, which is a symbolic expression, during the execution of a Theano function. If the condition is met, then the program will drop into the Python debugger (\verb|pdb|), and make available the values associated to a list of pre-defined \emph{monitored} variables. This is especially useful when something goes wrong during the training of a model, but only after a number of iterations, so it is not practical to log all values all the time.

        \subsubsection{Keeping the creation stack trace}
        \label{sec:new:debugging:stacktrace}
        
When a variable is created, part of the stack trace is recorded, in particular the line of the call that created it. For instance, if variable \verb|z| is created by calling \verb|z = a + b|, then the line where that expression is called is associated to \verb|z|. If evaluating that expression fails, for instance because \verb|a| and \verb|b| have incompatible shapes, then the error message will mention that file, line, and line number.

A challenge of that mechanism is that, when optimizations are applied, the replacement variables are not created at the same place as the ones they replace (or that ``correspond'' to them in a more general sense). In fact, they are created inside the optimization, so no stack trace is associated to them. For instance, if the expression above is optimized to move \verb|a| and \verb|b| to a GPU, and \verb|z| gets replaced by \verb|host_from_gpu(gpu_z)| where \verb|gpu_z = gpu_add(gpu_a, gpu_b)|, then the replacement for \verb|z| can easily retain the original stack trace, but \verb|gpu_z| would not.

To improve this feature, we are currently in the process of going through all optimizations, so that they assign the creation stack trace of the original variable (or variables) to the ``corresponding'' or equivalent one when they create replacements or new intermediate variables.

\section{Benchmarks}
\label{sec:benchmarks}


This section aims at giving a sense of the performance might expect from Theano against some of its largest competitors among machine learning research software, on different kinds of models. We used publicly-available software to compare against, when possible. We have made some of the benchmarking code public as well already, and will try to provide the remaining code as well in the future.

The goal of having more extensive benchmarks, on a wider variety of models and frameworks, is more easily attained by online projects, that can provide a picture more up-to-date. Among these projects, we can cite convnet-benchmarks,\footnote{\url{https://github.com/soumith/convnet-benchmarks/}} rnn-benchmarks,\footnote{\url{https://github.com/glample/rnn-benchmarks}} and hopefully DeepMark\footnote{\url{https://github.com/DeepMark/deepmark}} in the future.

We benchmarked Theano against Torch and TensorFlow (Section~\ref{sec:benchmark:setup}), on three kinds of popular machine learning models: convolutional networks (Section~\ref{sec:benchmark:convnet}), recurrent neural networks (Section~\ref{sec:benchmark:rnn}), and recurrent neural networks for sequence-to-sequence mapping (Section~\ref{sec:benchmark:seq2seq}). Finally, we show how the computation speed scales when using multiple GPUs with Platoon (Section~\ref{sec:benchmark:platoon}).

    \subsection{Setup}
    \label{sec:benchmark:setup}
All the benchmarks were run on a NVIDIA Digits DevBox, with 4 Titan X GPUs, and a Core i7-5930K CPU. All the benchmarks except for data-parallelism were run on only one GPU, which was not the one used for running the X server (using \verb|CUDA_VISIBLE_DEVICES|). We used Cuda 7.5.17, with cuDNN v4 (version 4007), and data type float32, for all frameworks and all experiments.

The compared software were installed as follow:
\begin{itemize}
    \item Theano was installed from the development version, at commit \verb|1bd371c|. The following configuration flags were used: \verb|floatX=float32|, \verb|lib.cnmem=0.45|, \verb|device=gpu0|, \verb|optimizer_including=unsafe|, \verb|dnn.conv.algo_fwd=time_once|, \verb|dnn.conv.algo_bwd_filter=time_once|, \verb|dnn.conv.algo_bwd_data=time_once|. For fast\_compile experiments, the additional option \verb|optimizer=fast_compile| was provided.
    \item TensorFlow 0.8 was installed from the binary package.
    \item Torch7 was installed from \url{https://github.com/torch/distro} at commit \verb|ffffc39|.
\end{itemize}

    \subsection{Convolutional networks}
    \label{sec:benchmark:convnet}

We measure the performance of four different convolutional models, that have been successfully used on the Imagenet dataset:
\begin{itemize}
    \item AlexNet, the one-column variant from~\cite{DBLP:journals/corr/Krizhevsky14}, with a batch size of 128;
    \item OverFeat, the \emph{fast} variant from~\cite{DBLP:journals/corr/SermanetEZMFL13}, with a batch size of 128;
    \item VGG, also known as OxfordNet, model A~\cite{DBLP:journals/corr/SimonyanZ14a}, with a batch size of 64;
    \item GoogLeNet V1~\cite{googlenet}, with a batch size of 128.
\end{itemize}
We used the code from \url{https://github.com/soumith/convnet-benchmarks} at commit \verb|84b5bb1| for Theano, Torch, and TensorFlow. We report the processing time per minibatch, for the forward and the backward pass.

\begin{figure}[htp]
\centering
\includegraphics[]{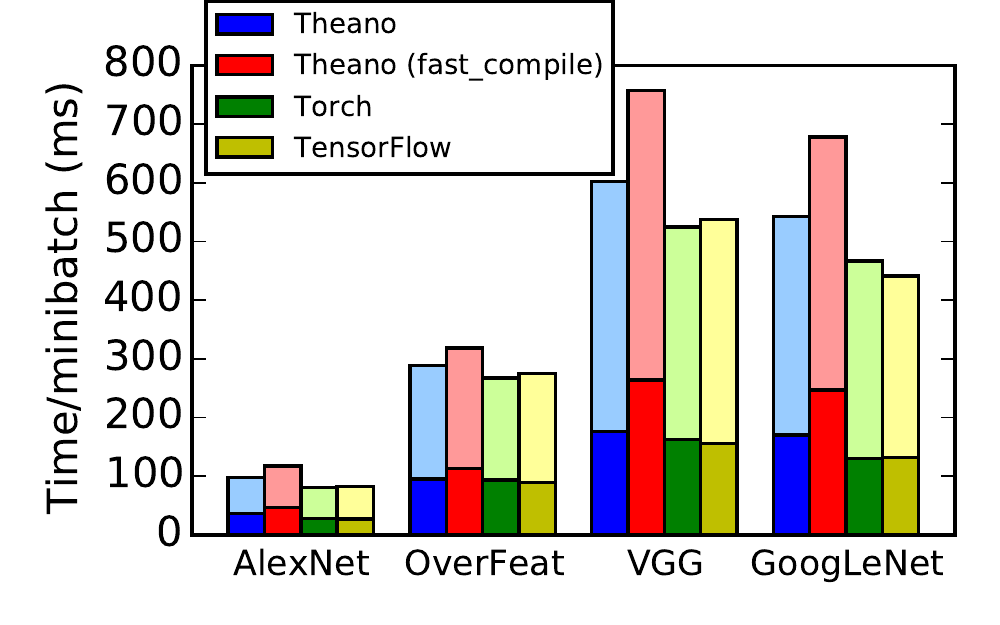}
\caption{Processing time for convolutional networks on Imagenet (milliseconds per batch, lower is better). Dark colors show forward computation time, pale colors show backward time.}
\label{fig:benchconvnet}
\end{figure}

The results, presented in Figure~\ref{fig:benchconvnet}, show that Theano is slightly slower than Torch and TensorFlow, but the performance is comparable, both for the forward and the backward passes. Furthermore, using the \verb|fast_compile| optimizer shows a slow-down between 10\% and 25\% only, which is a reasonable trade-off when developing or exploring a new model.

    \subsection{Recurrent neural networks: LSTM on Penn Treebank}
    \label{sec:benchmark:rnn}

To showcase recurrent network models, we benchmarked variants of the LSTM model applied to the Penn Treebank dataset described in~\cite{DBLP:journals/corr/ZarembaSV14}. We compared:
\begin{itemize}
    \item the Torch implementation available at \url{https://github.com/wojzaremba/lstm};
    \item the TensorFlow implementation showcased at \url{https://www.tensorflow.org/versions/r0.8/tutorials/recurrent/};\footnote{Code at \url{https://github.com/tensorflow/tensorflow/tree/master/tensorflow/models/rnn/ptb}}
and
    \item the Theano implementation available at \url{https://github.com/caglar/rnn_benchmarks}.
\end{itemize}

We measured words per second during training, and report results on the following models:
\begin{itemize}
    \item Small: Single Layer, 200 hidden units, sequence length: 20;
    \item Medium: Single Layer, 600 hidden units, sequence length: 40;
    \item Large: Two Layers, 650 hidden units each, sequence length: 50.
\end{itemize}
All three models used dropout on non-recurrent connections during training, following~\cite{DBLP:journals/corr/ZarembaSV14}. The batch size was set to 20.

\begin{figure}[htp]
\centering
\includegraphics[]{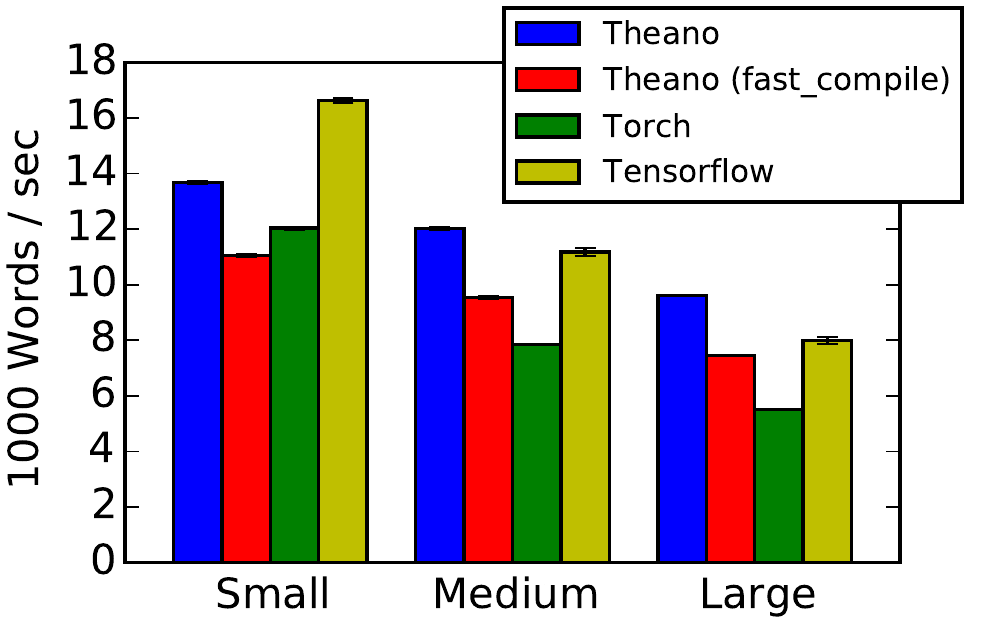}
\caption{Processing speed for different LSTM models on the Penn Treebank data set (words per second, higher is better).}
\label{fig:benchlstm}
\end{figure}

Figure~\ref{fig:benchlstm} shows that Theano comes second behind TensorFlow for the small model, but is slightly faster on the medium and large model. Torch was slower than Theano on all three models, and perhaps more surprisingly, slower than the fast\_compile version of Theano on the two larger models.

    \subsection{Sequence-to-sequence: Caption generation from video}
    \label{sec:benchmark:seq2seq}

In this section, we use the sequence-to-sequence mapping model from~\cite{yao2015capgenvid}. The input is a series of video frames and the output is a one-sentence English description of the input. Each input video frame is preprocessed by a GoogLeNet that was pre-trained for classification on ImageNet. The representation of the frame is thus a 1024 vector. The entire input is therefore represented by (M, F, 1024) where M is the minibatch size, and F is the number of frames. The output size is (M, L), where M is the minibatch size and L the sentence length (padding is used within a minibatch to ensure the same length, but different minibatches could have different L). Specifically, the model is written as $P(S|V)$, an LSTM on the sentence $S$, conditioned on the video $V$. $V$ is a weighted sum of frames representations.

The original code for~\cite{yao2015capgenvid} is available at \url{https://github.com/yaoli/arctic-capgen-vid}. We used simplified versions, in Theano and TensorFlow, instrumented for profiling, which will be made public in the future. There was no publicly available implementation in Torch. Theano with fast\_compile could not run because it was requiring too much memory. We report the processing time per minibatch, for the forward and backward passes, using three different batch sizes.

\begin{figure}[htp]
\centering
\includegraphics[]{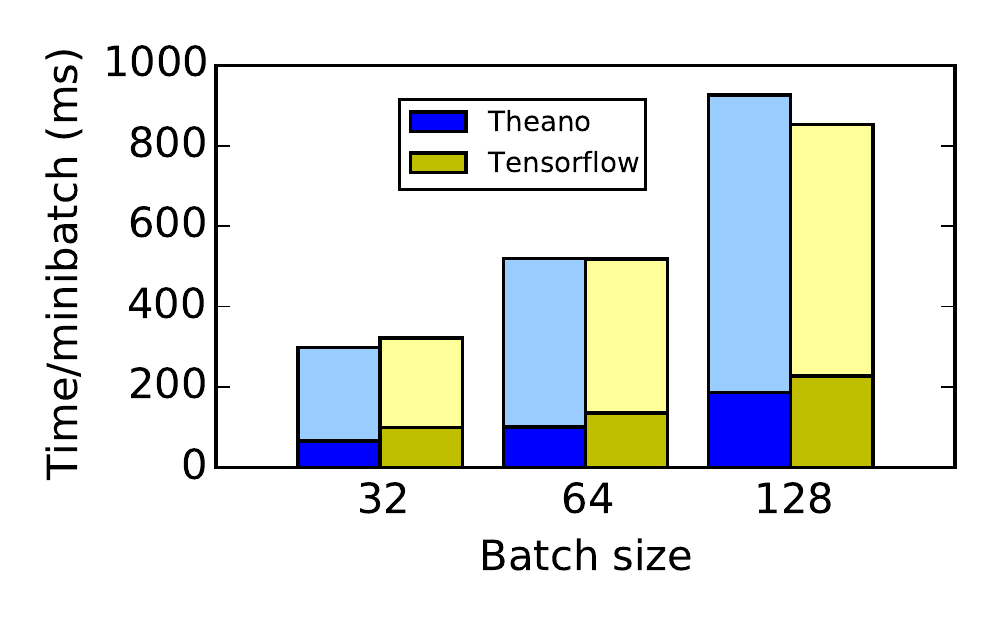}
\caption{Processing time for generating word sequences from video representations (milliseconds per batch, lower is better). Dark colors show forward computation time, pale colors show backward time.}
\label{fig:benchvideo}
\end{figure}

Figure~\ref{fig:benchvideo} shows a small advantage to Theano for the forward pass, but a disadvantage for the backward pass. The total time was comparable overall, with Theano being slightly faster on smaller batches, and TensorFlow being faster on larger ones. As expected, the time per minibatch grows slower than the minibatch size, because the potential for parallel computation is greater with larger batches.

    \subsection{Data parallelism for LSTM}
    \label{sec:benchmark:platoon}

We re-use the models from Section~\ref{sec:benchmark:rnn}, this time using Platoon to train on multiple GPUs on the same machine, using ASGD. We report results for 2 GPUs (using devices \verb|gpu1| and \verb|gpu2|) and 4 GPUs, compared against the results on 1 GPU obtained without Platoon and reported in Section~\ref{sec:benchmark:rnn}. We measured the overall processing speed (words per second) during training when synchronizing the models after every minibatch, and when synchronizing only every 100 batches.
The benchmarking code using Platoon will be made public soon.

\begin{figure}[htp]
\centering
\includegraphics[]{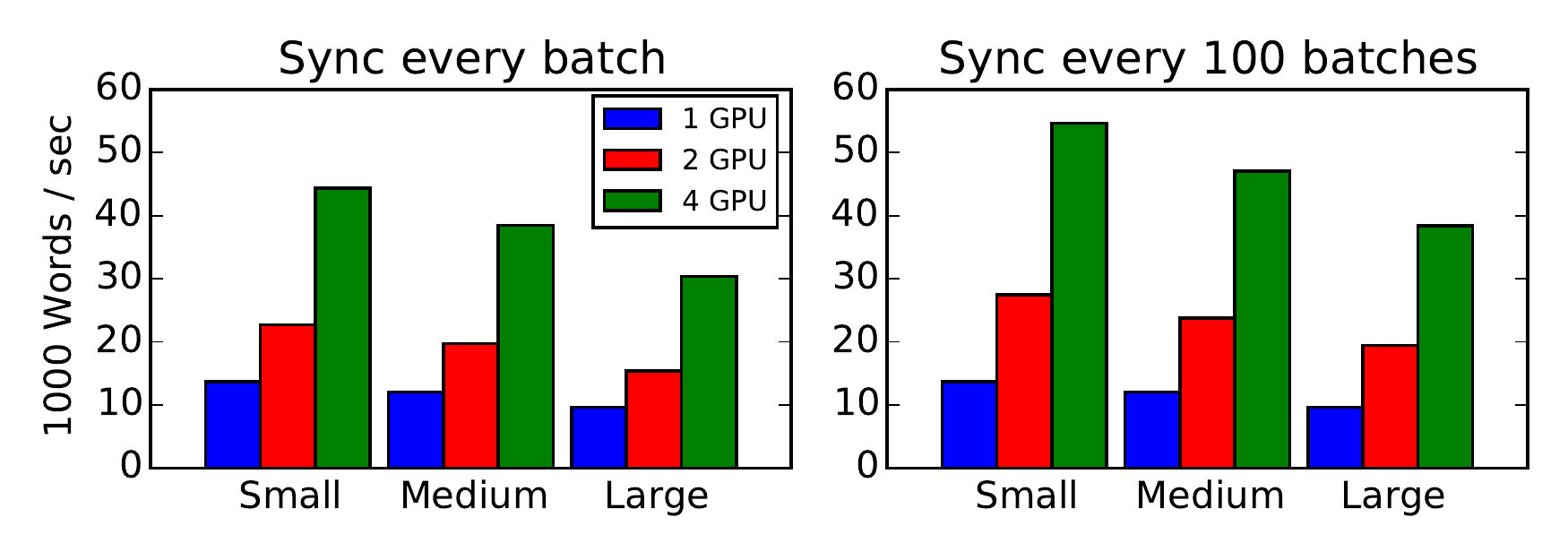}
\caption{Processing speed using multiple GPUs with Platoon on different LSTM models, synchronizing after each batch (left) and every 100 batches (right) (1000 words per second, higher is better).}
\label{fig:benchplatoon}
\end{figure}

Figure~\ref{fig:benchplatoon} shows a consistent increase in processing speed when adding more GPUs. As can be seen on the left, communication and synchronization overhead make that scaling sub-linear when synchronizing after every single batch, we found a speed-up between 1.6 and 1.7 for 2 GPUs and around 3.2 for 4 GPUs across all three models. Synchronizing only every 100 batches, on the right, brings the computation speed-up close to the theoretical optimum, at 2 for 2 GPUs and between 3.9 and 4 for 4 GPUs.

\section{Limitations and challenges}
\label{sec:challenges}

Despite the progress made in recent years and our best efforts, there remain some limitations or shortcomings in Theano. Some of these issues have been addressed by competing frameworks mentioned in Section~\ref{sec:main:comparison}, and by other projects like CGT (Computation Graph Toolkit).\footnote{\url{http://rll.berkeley.edu/cgt/}}

    \subsection{Limitations from Python}
    \label{sec:challenges:python}

Since Theano uses Python as its core language, and uses NumPy arrays and other Python objects to store values, it is affected by Python's limitations. The main one is the Python GIL, that limits concurrent execution of threads. We have seen that it is possible to make single-threaded execution fast by compiling binary modules that are then loaded in Python (Sections~\ref{sec:main:compiler:c_code} and~\ref{sec:main:library}), and it would also be possible to release the GIL during the execution of these functions. However, the GIL has to be acquired again each time references to Python objects are added or removed, when using the C API of Python and NumPy. Since the execution of such functions is usually quite short, most threads would spend their time waiting for the lock instead of performing actual computation.

Since Python has a concept of threads and expects to be in charge of threading, it is also not possible to launch different, independent Python interpreters in different threads of the same process, as is possible with Lua for instance.

To avoid that issue, we could use a different n-dimensional array structure, that is accessible directly from C++ without actually being a Python object, like the one libgpuarray provides on the GPU. It would require Theano to explicitly manage memory allocation and deallocation, in a thread-safe way. It would also require to rewrite all the C++ and CUDA code for existing Ops, so that they use a different interface for reading their input data and writing their output data. Finally, it could make it harder to create new Ops by integrating existing Python code.

    \subsection{Graph optimization time}
    \label{sec:challenges:opttime}

The execution time of the graph optimization phase is not scaling well with graph size. Currently, it is scaling supra-linearly relative to the number of nodes. One issue is that some groups of local optimizations try to apply over and over, until none of them can be applied any more, and the graph stops changing. In practice, it can force a number of passes through the whole graph that becomes bigger for bigger graphs (the chances of some local optimization applying somewhere are higher).

An option would be to completely reorganize the existing optimizations so that they are more lightweight, and can be applied in a fixed number of passes through the graph. It could be possible, for instance, to use a one-pass or two-pass optimization phase, like CGT does. Doing that without any regressions in the stability optimizations could be a large-scale project.

    \subsection{Code compilation time}
    \label{sec:challenges:compilation}

Currently, the same Theano Op can generate a large quantity of different C++ or CUDA modules, depending on its properties at compile time, such as the data type of inputs and outputs, whether it will run in place, and other flags determining its behaviour. Compiling and loading those modules can take time and add a load on the file system.

To alleviate those issues, it would be possible in most cases to pass that information dynamically at runtime, instead of hard-coding it in the generated code. This approach is already being used in the new back-end to specify which GPU should be used for the execution of a particular Apply node, but it could be generalized.

    \subsection{Loops and control-flow structures}
    \label{sec:challenges:loops}

Using Scan for loops, and the \verb|ifelse| lazy Op for conditionals, has proven a useful way of expressing control-flow operations. However, with an increasing need for more flexibility (attention mechanisms, nested loops, recursive loops, changes in shape between iterations of the same loop), we may need a more principled way of expressing these structures.

One appealing way would be to use \emph{switch} and \emph{merge} Apply nodes in the computation graph, like in a dataflow graph~\cite{arvind1986dataflow}. This is the approach taken by TensorFlow~\cite{tensorflow2015-whitepaper} for symbolic loops. This would require adding support for cycles in the computation graph in these circumstances, extending the runtime to be able to recompute values inside the loop, and rewriting all the graph optimizations currently existing for Scan, including the ones limiting memory consumption.

    \subsection{Multi-node parallelism}
    \label{sec:challenges:multinode}

Scaling model execution and training to multiple machines is outside of the scope of Theano's core, but additional packages could be developed to interface with Theano, in the same way Platoon does for multiple GPUs in a single node. In fact, tools like parameter servers and coordinators do not have to be specific to Theano, and could be common to different frameworks.

    \subsection{Improving memory usage}
    \label{sec:challenges:memory}

Given the limited availability of on-board GPU memory, memory consumption is often a bottleneck for training machine learning algorithms. This can limit the size and modelling power of trainable models, and make the processing power of GPUs under-used, for instance when batch sizes have to be reduced.
In addition to storing intermediate values in a lower-precision format (for instance, storing data as float16 is supported in Theano's new GPU back-end), different options could be explored and combined:
\begin{itemize}
    \item Change the order of execution of computations, so the peak memory usage is reduced. This can be done statically before the function is executed, or dynamically, for instance by detecting that memory is insufficient and waiting for some other computation to finish and free intermediate values.
    \item Move intermediate values to the main (CPU) memory, or to another GPU's memory, if it is not needed for a while, and transfer it back before it is used again. This method has been successfully implemented by~\cite{2016arXiv160208124R}.
    \item Free intermediate values, and recompute them when they are needed again. This approach has been used in~\cite{2016arXiv160406174C}, and can be especially useful for fast operations that have large outputs.
\end{itemize}

    \subsection{The future of gradient-based computation frameworks}
    \label{sec:challenges:future}

Tools like Theano and TensorFlow are compilers for mathematical expressions, in that they require the code (or computation graph) to be defined first, and then executed. On the other hand, Torch works more like an interpreter: the computation is done as soon as the expression is called. It could be interesting to explore how to apply JIT (just-in-time) compiler ideas to the computation graph, to combine the immediate response and flexibility of an interpreter (including using control flow statements like \verb|if|, \verb|for|, \verb|while|, from the language directly), and the performance gains of a compiler when an expression has to be evaluated multiple times.

Most machine-learning frameworks can now share efficient implementations of GPU kernels, such as the ones published by NVIDIA (cuDNN) and Nervana. Graph optimizations could be another component shared between projects, maybe through a common language to define computation graphs and such optimizations. It could be common to machine learning frameworks and computer algebra systems (CAS) such as SymPy~\cite{sympy} and SympyCore.\footnote{\url{https://github.com/pearu/sympycore}}

\section{Conclusion}

Theano pioneered ideas for efficient gradient-based computation that are now part of most mainstream machine-learning research libraries, for instance combining a high-level scripting language with highly-optimized computation kernels, especially using GPUs, symbolic computation graph, and symbolic differentiation. Some other features of Theano, like graph rewriting and optimizations, and automatic generation and compilation of kernels, are starting to become more widely used as well.

Continuous improvements have been made to Theano's functionality, usability, and performance, for instance wrapping libraries like cuDNN, and integrating ideas that have been successfully explored and implemented by other frameworks, like data parallelism and model parallelism for distributed computation.
Computation performance is on par with other major research software, like Torch and TensorFlow.

There are ways to improve Theano (and other frameworks as well) by taking inspiration from other machine learning software (sometimes more experimental). Longer-term improvements could be the result of collaborations with other fields, for instance CAS, and language and compiler design, in order to build a next generation of mathematical computation software.

\begin{acknowledgments}
We acknowledge the support of the following organizations for research funding and computing support: NSERC, Calcul Québec, Compute Canada, the Canada Research Chairs, and CIFAR.

Lijun Xue, Qianli Ma, Ziye Fan, and Christof Angermueller contributed to Theano through the Google Summer of Code program.


The authors would like to thank all the other committers to Theano:
Faruk Ahmed,
Diogo Moitinho de Almeida,
Hani Almousli,
Andrea,
Martin Andrews,
John Arevalo,
Martin Arjovsky,
Kai Arulkumaran,
Ben Athiwaratkun,
bbabeshkin,
Markus Beissinger,
Sebastian Berg,
Thierry Bertin-Mahieux,
Lucas Beyer,
Merlijn Blaauw,
Jörg Bornschein,
Ethan Buchman,
Bogdan Budescu,
Yaroslav Bulatov,
Liwei Cai,
Brian Cheung,
Claude Coulombe,
Frans Cronje,
Rolf van Dam,
Jonas Degrave,
Misha Denil,
Doug,
Zach Dwiel,
Ilya Dyachenko,
Douglas Eck,
Michael Eickenberg,
Amir Elaguizy,
eulerreich,
Marco Fagiani,
Raul Chandias Ferrari,
Abraham Flaxman,
Mike C. Fletcher,
Piotr Frankowski,
Geoffrey French,
Adithya Ganesh,
Dario Garcia,
Sergii Gavrylov,
Wojciech Głogowski,
Matthew Koichi Grimes,
gw0,
Christophe Van Gysel,
Yaroslav Halchenko,
Tim Head,
Hei,
Jonathan Ho,
Paul Hollensen,
Andre Georg Holzner,
Liang-Chi Hsieh,
Eric Hunsberger,
Jonathan J. Hunt,
Vlad Ionescu,
Andy Jiang,
jojolalpin,
joncrall,
Yi-Lin Juang,
Vik Kamath,
Moslem Kazemi,
Kevin Keraudren,
Robert Kern,
Marius Killinger,
Taesup Kim,
Jey Kottalam,
Stefan Krastanov,
Gokula Krishnan,
Matthias Kümmerer,
Kosuke Kusano,
Micky Latowicki,
Eric Laufer,
Sergei Lebedev,
Rémy Léone,
Wei Li,
Peng Liu,
Jakob Lombacher,
Gilles Louppe,
Jan-Matthis Lückmann,
Michael I. Mandel,
Daniel Maturana,
Sergey Matyunin,
Madison May,
Ben McCann,
Clay McLeod,
Thomas Mesnard,
Grégoire Mesnil,
Luke Metz,
Kyle Meyer,
Marco De Nadai,
Anchit Navelkar,
Alassane Ndiaye,
Huy Nguyen,
Michael Opitz,
Johannes Otterbach,
Wei Ouyang,
Daniil Pakhomov,
Seon-Wook Park,
Fábio Perez,
Steven Pigeon,
Nicolas Pinto,
Zach Ploskey,
Bhavishya Pohani,
Ben Poole,
Rahul,
Sirisha Rambhatla,
Kashif Rasul,
Julien Rebetez,
Marc-Antoine Rondeau,
Tim Salimans,
Adam Salvail,
Joao Felipe Santos,
Utkarsh Saxena,
Ludwig Schmidt-Hackenberg,
Ilan Schnell,
Hannes Schulz,
Anish Shah,
Saatvik Shah,
Shai,
Yurii Shevchuk,
Scott Sievert,
Søren Kaae Sønderby,
spotted1234,
Graham Taylor,
Texot,
theaverageguy,
Martin Thoma,
Carl Thomé,
Chiheb Trabelsi,
Matthew Trentacoste,
Christos Tsirigotis,
Karen Ullrich,
Prayag Verma,
Karel Vesely,
Mang Wang,
XterNalz,
Yu Yang,
yobibyte,
Jason Yosinski,
Lee Zamparo,
John Zedlewski,
Albert Zeyer,
and
ziyuang.

\end{acknowledgments}

\bibliography{theano2016}

\end{document}